\documentstyle[eqsecnum,aps]{revtex}

\begin{document}
\draft
\title{Einstein Manifolds, Abelian Instantons,\\
Bundle Reduction, and the Cosmological Constant}
\author{Chopin Soo\cite{byline1}\\
}
\address{
Physics Department,
National Cheng Kung University,\\
Tainan, Taiwan 70101, Taiwan.
}
\maketitle
\begin{abstract}
  The anti-self-dual projection of the spin connections of certain four-dimensional
  Einstein manifolds can be Abelian in nature. These configurations signify
  bundle reductions. By a theorem of Kobayashi and Nomizu such a
  process is predicated on the existence of a covariantly constant field.
It turns out that even without fundamental Higgs fields and other
physical matter, gravitational self-interactions can generate this
mechanism if the cosmological constant is non-vanishing. This
article identifies the order parameter, and clarifies how these
Abelian instanton solutions are associated with a Higgs triplet
which causes the bundle reduction from SO(3) gauge group to U(1).

\end{abstract}

\widetext

\section{Introduction}

   Recent Type Ia Supernovae observations strongly support
   a non-vanishing and positive cosmological constant\cite{Perlmutter}.
   In four-dimensions many exact and interesting solutions of Einstein's field equations
   $R_{\mu\nu} = \lambda g_{\mu\nu}$ with
   cosmological constant, $\lambda$, are known.
   With a non-vanishing cosmological constant the (anti)self-dual part of the
   spin connections of certain Einstein manifolds can be Abelian.
   Such a solution of Einstein's equations signify a bundle reduction.
   For this to happen a theorem of Kobayashi and Nomizu
   demands the existence of a covariantly constant field
   on an associated bundle\cite{Kobayashi-Nomizu}.
   The relation of this field to the cosmological constant
   in the bundle reduction process is clarified in this article.
   Although Kobayashi and Nomizu's theorem requires such fields
   for all bundle reductions,
   it sheds no further light on their precise nature.
   Indeed it remains to be seen whether current particle physics searches for
   Higgs fields responsible for symmetry breaking in the
   standard model will encounter fundamental or composite objects.
   Kobayashi and Nomizu's theorem however seems to guarantee their
   existence in the description of bundle reduction processes. It
   is therefore remarkable Abelian instantons discussed in
   this article demonstrate that with a non-vanishing cosmological constant
   gravitational self-interactions can generate the
   mechanism of bundle reduction dynamically, despite the absence of
   additional fundamental Higgs fields and other physical matter.

   After an introduction on self and anti-self-dual
   decompositions of the Riemannian curvature in the next section a
   couple of explicit Abelian instantons configurations are
   recalled in Section III. The theorem of Kobayashi and Nomizu is
   recounted in Section IV, together with its specialization to
   bundle reduction of the $SO(3)$ gauge group to $U(1)$. The role of
   the cosmological constant is clarified in Section V, and the article
   ends with a note on Abelian instantons and their relation to
   K{\"{a}}hler-Einstein manifolds in Section VI.

\section{Self and anti-self-dual decompositions of the Riemannian curvature}

   In contradistinction to $SO(3,1)$ for Lorentzian
   signature, the local Euclidean symmetry of the theory
   is gauged by an $SO(4)$ spin connection in the case of Euclidean
   signature discussed in this article. See, for instance, Ref.\cite{Hawking}
   on the importance of Euclidean instanton solutions in semi-classical
   and quantum gravity.
   Since $SO(4) = [SU(2)\times SU(2]/Z_2$,
   the self and anti-self-dual parts of the spin connection are
   generically $SO(3)$ in value (in the Lorentzian instance $SO(3,1)$
   is isomorphic to $SO(3,C)$, and self and anti-self-dual projections are related
   by complex conjugation). It is both natural and useful to analyze the properties
   of four-dimensional Einstein manifolds in
   terms of these projections of the spin connections.
   Four dimensions also has the peculiarity of allowing a decomposition of the
   Riemann curvature two-form into components taking values in the
   $(\pm 1)$ eigenspaces, $\Lambda_2^\pm$, of the Hodge duality operator $*$.
   To be precise, the bases for $\Lambda^{\pm}_2$ can regarded to
   be the self and anti-self-dual two-forms
   $\Sigma^\pm_a \equiv \pm e_0 \wedge e_a
   +\frac{1}{2}\epsilon_{0abc} e^b \wedge e^c$ with $e_A$ denoting
   the vierbein. Explicitly $* \Sigma^\pm = \pm
  \Sigma^\pm$. The conventions are that upper case Latin Euclidean Lorentz
  indices take values from 0 to 3, while lower case indices run from 1 to 3.
Spacetime components are represented by Greek indices.

   The Riemann-Christofel curvature tensor $R_{ABCD}$ has four indices
   (anti-symmetric in pairs), and can be dualized both on the left
   and on the right. These can be thought of as
   internal and external duality transformations if one considers the
   Riemann curvature two-form $\frac{1}{2}R_{AB\mu\nu}dx^\mu \wedge dx^\nu$; or as
   acting wholly on internal Lorentz indices if one considers the curvature
   components of $\frac{1}{2}R_{ABCD}e^C \wedge e^D$ expressed in vierbein basis. For convenience
   this latter point of view is adopted here. Given a non-degenerate vierbein,
   these concepts are interchangeable.
   The duality operation will be denoted by a tilde over the
   indices, so $R_{{\widetilde{AB}}CD} \equiv
   \frac{1}{2}\epsilon_{ABEF}R^{EF}\,_{CD}$ implies dualizing on the left and
   $R_{AB{\widetilde{CD}}} \equiv
   \frac{1}{2}\epsilon_{CDEF}R_{AB}\,^{EF}$ on the right.
   In the decomposition into self and
   anti-self-dual projections this mapping $\Lambda^\pm_2 \mapsto \Lambda^\pm_2$
   can be viewed as a $6 \times 6$ matrix which assumes the form
   $\left[{\matrix{S^+ & C^+ \cr C^- & S^-}}\right]$\cite{AHS,EH}.
   In terms of curvature components, the elements of the $3\times 3$ matrix
   $S^+_{ab}$ are
\begin{equation}
S^+_{ab} \equiv + \frac{1}{2}(R_{0a0b} + R_{0a\widetilde{0b}}) +
\frac{1}{2}(R_{\widetilde{0a}0b} +
R_{\widetilde{0a}\widetilde{0b}}),
\end{equation}
while $S^-$ and $C^\pm$ are defined by replacing the signs in the
definition of $S^+$ above following the prescription
\begin{eqnarray}
  S^+ \sim(+,+,+,+)  &,& \qquad S^- \sim(+,-,-,-), \cr
  \nonumber\\
  C^+ \sim(+,-,+,-) &,& \qquad C^- \sim(+,+,-,+).
\end{eqnarray}
Consequently it can be checked that $S^+(S^-)$ is self-dual
(anti-self-dual) with respect to both left and right duality
operations, while $C^+(C^-)$ is self-dual (anti-self-dual) under
left duality and anti-self-dual (self-dual) under right duality
operations. Furthermore it is known\cite{AHS} that a metric
satisfies Einstein's equations if and only if $C^\pm$ vanishes
i.e. the $6\times 6$ matrix above assumes the block diagonal
form, and its trace equals $- 2\lambda$. On-shell, the doubly
self and anti-self-dual $S^\pm$ correspond precisely to the
curvature components of the Ashtekar connections\cite{Ash}.

While not all Einstein manifolds possess (anti)self-dual
Riemann-Christofel curvatures, it follows from the block diagonal
requirement that a manifold is nevertheless Einstein only if the
(anti)self-dual part of its spin connection is also doubly
(anti)self-dual. So when expressed in terms of (anti)self-dual
spin connections $A^\pm_a = \omega_{0a}
 \pm \frac{1}{2}\epsilon_{0abc} \omega^{bc}$,
 {\it all} Einstein manifolds are (anti)instantons in the sense
$*F^\pm_a = \pm F^\pm_a$, with $F^\pm_a \equiv d A^\pm_a
+\frac{1}{2}\epsilon_{0abc} A^{\pm b}\wedge A^{\pm c})$
\cite{csaction}. Many known exact solutions also have finite
Einstein-Hilbert action.

 The covariant Samuel-Jacobson-Smolin actions\cite{SJS} which capture the
canonical variables and constraints of Ashtekar can be written as
\begin{equation}
   {\cal A}^\pm = {1\over{16\pi G}}\int (2 F^{\pm a} \wedge \Sigma^\pm_a
   + {\lambda \over 3}\Sigma^{\pm a} \wedge \Sigma^\pm_a).
\end{equation}
These are just the self and anti-self-dual projections of the
Einstein-Hilbert-Palatini action, plus the usual cosmological term
written in terms of $\Sigma^\pm$. Remarkably {\it either} the $+$
{\it or} $-$ action is all one needs to reproduce equations of
motion equivalent to Einstein's theory. Note that $\pm$ variables
are interchanged under an orientation reversal such as under
parity transformation. In this article $-$ i.e. anti-self-dual
variables and the corresponding action will be adopted.

The equations of motion which follow from variations of $A^-_a$,
$e_0$ and $e_a$ are respectively
\begin{eqnarray}
 D_{A^-}\Sigma^{-a} &=&0,\cr
 \nonumber\\
 2F^-_a \wedge e^a +\frac{\lambda}{3}\epsilon_{0abc}e^a\wedge
 e^b\wedge e^c &=&0,\cr
 \nonumber\\
2F^-_a \wedge e_0 + 2\epsilon_{0abc}e^b \wedge F^{-c} + \lambda
\epsilon_{0abc}e^0\wedge e^b\wedge e^c &=&0.
\end{eqnarray}
The first equation is solved by $A^-_a = \omega_{0a}-
\frac{1}{2}\epsilon_{0abc}\omega^{bc}$, where $\omega_{AB}$ is the
torsionless spin connection respecting $de_A + \omega_{AB}\wedge
e^B =0$. Since $F^-_a$ is anti-self-dual and $\Sigma^-_a\wedge
\Sigma^-_b = -2\delta_{ab}(*1)$, it is possible to expand $F^-_a$
as
\begin{equation}
  F^-_a = S^-_{ab}\Sigma^{-b}, \label{Ss}
\end{equation}
where $S^-_{ab}= -\frac{1}{2}(F^-_a\wedge \Sigma^-_b)$ is
constrained by
\begin{equation}
 S^-_{ab}=S^-_{ba} \qquad {\rm and} \quad Tr (S^-)= -\lambda
\end{equation}
to satisfy the rest of the equations of motion \cite{cps1,cps2}.
On-shell, $S^-_{ab}$ is identical to the $3\times 3$ matrix with
the same notation mentioned previously in Eq.(2.2) (see also
Eq.(A5) of the Appendix). The Appendix contains a brief proof of
the equivalence of this description to Einstein's field
equations. There have been suggestions to arrive at a
metric-independent description of General Relativity through
$\Sigma^-_{a} = (S^-)_{ab}^{-1}F^{-b}$\cite{cdj}. However, the
Abelian solutions addressed in this article demonstrate that this
cannot be done in general, since for these configurations $S^-
={\rm diag}(0,0,-\lambda)$ is non-invertible. On the other hand
Eq.(\ref{Ss}) remains valid for all Einstein manifolds with
non-degenerate vierbeins.

\section{Abelian Instantons: Explicit examples}

The examples can be expressed in terms of four-dimensional polar
coordinates $(R,\theta, \phi,\psi)$ with
  $R^2 \equiv \sum^{K=3}_{K=0} x^K x^K, 0\leq \theta < \pi,
  0\leq \phi < 2\pi, 0\leq \psi < 4\pi$; and
  \begin{eqnarray}
    \Theta_1 & = & \frac{1}{2}(\sin\psi d\theta -\sin\theta\cos\psi
    d\phi),\cr
    \nonumber\\
   \Theta_2 & = & \frac{1}{2}(-\cos\psi d\theta -\sin\theta\sin\psi
    d\phi),\cr
    \nonumber\\
    \Theta_3 & = & \frac{1}{2}( d\psi + \cos\theta d\phi).
  \end{eqnarray}
$\Theta_a$ satisfy $d\Theta_a = \epsilon_{abc} \Theta^b \wedge
\Theta^c$, and $-2\Theta_a$ are the Maurer-Cartan 1-forms on
$S^3$.

(i) Consider first the complex projective space $\overline{CP}_2$
and the well-known Fubini-Study metric
\begin{equation}
ds^2 = {{(dR)^2}\over{(1+ \frac{\lambda}{6}R^2)^2}} +
{{(R\Theta_1)^2}\over{(1+ \frac{\lambda}{6}R^2)}} +
{{(R\Theta_2)^2}\over{(1+ \frac{\lambda}{6}R^2)}} +
{{(R\Theta_3)^2}\over{(1+ \frac{\lambda}{6}R^2)^2}}
\end{equation}
The equation of motion $ D_{A^-}\Sigma^-_a =0$ implies $A^-_a =
\omega_{0a}- \frac{1}{2}\epsilon_{abc}\omega^{bc}$. Choosing the
vierbein as $e_A =\{{{dR}\over{(1 +\frac{\lambda}{6}R^2)}},
-{{R\Theta_1}\over{(1+ \frac{\lambda}{6}R^2)}^{1\over 2}},
-{{R\Theta_2}\over{(1+ \frac{\lambda}{6}R^2)}^{1\over 2}},
-{{R\Theta_3}\over{(1+ \frac{\lambda}{6}R^2)}}\}$ leads to
\begin{equation}
  A^-_1 = A^-_2 =0, \quad {\rm and} \quad A^-_3 = -{{\lambda R^2 \Theta_3}\over{
  2(1+ \frac{\lambda}{6}R^2)}},
\end{equation}
giving
\begin{equation}
  F^-_1 = F^-_2 =0, \quad {\rm and} \quad F^-_3 = dA^-_3
  =-\lambda\Sigma^-_3,
\end{equation}
i.e. $F^-_a = S^-_{ab} \Sigma^{-b}$ with $S^-_{ab}= {\rm
diag}(0,0,\lambda)$. The configuration is explicitly Abelian,
moreover $*F^-_a = -F^-_a$. \hfil

(ii)As the next example, take the famous Eguchi-Hanson manifold
$\overline{EH}$ with metric
\begin{equation}
ds^2 = [1-(a/R)^4 -\frac{\lambda}{6}R^2]^{-1}(dR)^2 +
(R\Theta_1)^2 + (R\Theta_2)^2 + [1-(a/R)^4
-\frac{\lambda}{6}R^2](R\Theta_3)^2.
\end{equation}
Adopting $e_A = \{[1-(a/R)^4 -\frac{\lambda}{6}R^2]^{-{1\over
2}}dR, -R\Theta_1, -R\Theta_2, -[1-(a/R)^4
-\frac{\lambda}{6}R^2]^{1\over 2}R\Theta_3\}$ results in
\begin{equation}
  A^-_1 = A^-_2 =0, \quad {\rm and} \quad A^-_3 = -{\lambda \over 2}R^2
  \Theta_3,
\end{equation}
and
\begin{equation}
  F^-_1 = F^-_2 =0, \quad {\rm and} \quad F^-_3 = -\lambda\Sigma^-_3,
\end{equation}
Again $F^-_a = S^-_{ab} \Sigma^{-b}$ with $S^-_{ab}= {\rm
diag}(0,0,\lambda)$ is explicitly Abelian and anti-self-dual with
respect to $*$.

It is a matter of convention whether a particular vierbein
orientation is associated to $M$ or $\overline{M}$. However it
should be mentioned that these projections of the spin
connections are orientation dependent. For instance, an
orientation reversal under parity transformation ($e_0 \mapsto
e_0, e_a \mapsto -e_a$) can produce non-Abelian configurations.
This corresponds to looking at $S^+$ instead. The full Riemannian
curvature of Einstein manifolds is contained in $S^+$ together
with $S^-$. If one of the projections is an Abelian configuration,
then by an appropriate choice of orientation it suffices to
consider the case of $S^-$ for which this is true. Configurations
with opposite orientations are not however not related by general
coordinate and gauge transformations if they have non-vanishing
invariants which are odd under orientation reversal. This is true
of the two examples given above. They possess non-vanishing
values of the signature invariant \cite{EH}. Further discussions
on these and related issues for Ashtekar connections can be found
in Ref.\cite{cps2}.

\section{A theorem of Kobayashi and Nomizu on bundle reduction, and the Higgs triplet
for $SO(3)$}

 A powerful theorem of Kobayashi and Nomizu\cite{Kobayashi-Nomizu} asserts
 that bundle reduction is predicated on the existence of a covariantly constant field.
 Proposition 7.4 of Ref.\cite{Kobayashi-Nomizu} repeated verbatim is:

\it{Let $P(M,G)$ be a principle fiber bundle and $E(M,G/H,G,P)$
the associated bundle with standard fiber $G/H$, where $H$ is a
closed subgroup of G. Let $\sigma: M \mapsto E$ be a cross
section and $Q(M,H)$ the reduced subbundle of P(M,G)
corresponding to $\sigma$. Then a connection $\Gamma$ is
reducible to a connection $\Gamma'$ if and only if $\sigma$ is
parallel with respect to $\Gamma$.}

\rm
 The theorem is, remarkably, contemporaneous with, if not earlier than, seminal
 investigations of symmetry breaking in gauge field theories in particle
 physics in the 1960s\cite{Nambuetal}. It also implies, as we shall see,
 that the bundle reduction is a gauge and diffeomorphism invariant
 concept. So the Abelian nature of these Einstein manifolds is physically meaningful
 in General Relativity.

 In the previous section Abelian
 instanton configurations were displayed, despite the gauge group of
 the $A^-_a$ connection being in general $SO(3)$ in value. By Kobayashi
 and Nomizu's proposition, a Nambu-Goldstone-Higgs field\cite{Nambuetal}
 $\sigma$ {\it must} be at work. Let us therefore proceed to demonstrate this
 below and in the next section. To wit, we specialize to $G = SO(3)$ and $H = SO(2)
= U(1)$. Furthermore let $\sigma$ correspond to a Higgs triplet
$\phi_a (a=1,2,3)$, $\Gamma$ to the connection $A^-_{\mu
a}dx^\mu$, and $\Gamma'$ to
 $A^-_{\mu a}\widehat{\phi}^a dx^\mu$ with $\widehat{\phi}^a
 \equiv \phi^a/\sqrt{\phi_b\phi^b}$.
  The clause ``{\it parallel with respect to} $\Gamma$"
 in the above proposition is now the same as
\begin{equation}
D_{A^-}\phi_a = d\phi_a + \epsilon_{0abc} A^{-b}\wedge\phi^c = 0
\label{Higgs}
\end{equation}
This is a well-known condition characterizing ``symmetry
breaking"\cite{Goddard-Olive}. It is instructive to recall some
results which are relevant to symmetry breaking of $SO(3)$ to
$U(1)$. Contracting $\phi^a$ with Eq.(\ref{Higgs}) results in
$d\phi^2 =d(\phi_a \phi^a) =0$. Thus the magnitude of
$\vec{\phi}$ is constant, and therefore the allowed values of
$\vec{\phi}$ lie on a 2-sphere of constant radius
$\sqrt{\phi_a\phi^a} = K$ in internal space. Regardless of what
direction $\vec{\phi}$ is pointing in internal space, it is
invariant with respect to Abelian $U(1)$ rotations about the
$\widehat{\vec{\phi}}$-axis. This symmetry forms the ``little
group" $H=U(1)$ of $\vec{\phi}$, while every other point on the
2-sphere is accessible by an $SO(3)/U(1)$ rotation. Thus the
allowed values of $\vec{\phi}$ are parametrized by $G/H$ with
$G=SO(3)$. After a little algebra contraction of
$\epsilon^{ade}\phi_d$ with Eq.(\ref{Higgs}) can be shown to
yield (for $K\neq 0$)
\begin{equation}
A^-_{a} = (A^-_{b}\widehat{\phi}^b)\widehat{\phi}_a +
\epsilon_{0abc}(d\widehat{\phi}^b)\widehat{\phi}^c. \label{Ugauge}
\end{equation}
This shows the decomposition of the $SO(3)$ connection into its
Higgs projection and a piece depending only on $\vec{\phi}$. We
may then perform in general local (spacetime-dependent) gauge
transformations to arrive at the U-gauge\cite{Goddard-Olive} in
which the Higgs field assumes $\phi^a = \phi^3\delta^a_3$. As a
result Eq.(\ref{Ugauge}) implies $A^-_{1} = A^-_{2} =0$, and
$A^-_{3} = (A^-_{3}\widehat{\phi}^3)\widehat{\phi}^3$ is the only
non-vanishing component. A reduction of the connection from
$A^-_{a}$ to Abelian $A^-_{3}$ has occurred, because $\vec{\phi}$
satisfies the conditions of the theorem as a consequence of
Eq.(\ref{Higgs}). It may appear that there could be obstructions,
especially for topologically nontrivial configurations, to the
existence of a globally defined U-gauge. However, nontrivial
configurations -for instance, 't Hooft-Polyakov monopoles
associated with symmetry breaking\cite{tHooft}- can be described
equally well within the U-gauge\cite{Goddard-Olive}. The reason
is although more than one coordinate patch is required, the
nontrivial topological information is captured by the transition
functions between the different patches. These transition
functions are therefore H-valued in the U-gauge, and so the
reduced bundle and connection are restricted to be H-valued
entities while retaining the topological information.

\section{Dynamical bundle reduction and order parameter
for the Abelian instantons}

In the previous section bundle reduction from $SO(3)$ to $U(1)$
was caused by a Higgs triplet obeying Eq.(\ref{Higgs}), in
accordance with the theorem of Kobayashi and Nomizu. Such a
mechanism is at work for Einstein manifolds with $A^-_a$ Abelian
instanton connections. Let us consider the order parameter
$S^-_{ab}$. As shown, for Abelian instantons it assumes the form
$S^-_{ab}={\rm diag}(0,0,-\lambda)$. This can be can translated
into manifestly gauge-covariant language. Suppose $\vec{\phi}$
has constant magnitude such that $\phi^2 \equiv \phi^a\phi_a
=\lambda > 0$. Then by an $SO(3)$ gauge rotation, the U-gauge
configuration $\phi_a = -\delta_a^3\sqrt{\lambda}$ can be
transformed into the generic Higgs triplet $\phi_a = O_{ab}\phi^3
\delta^b_3$ i.e. $\phi_a =(\phi_1,\phi_2,\phi_3)$, and $S^-_{ab}$
into $(OS^-O^T)_{ab}= -\phi_a \phi_b$. An explicit $SO(3)$ matrix
which achieves this purpose is
\begin{equation}
O = -\left[\matrix{
  {\phi_3\over{\sqrt{\phi^2_1 + \phi^2_3}}} &
  {{\phi_1\phi_2}\over{\sqrt{\phi^2(\phi^2_1 + \phi^2_3})}}
  &{{\phi_1}\over {\sqrt{\phi^2}}} \cr
  0 & {-\sqrt{\phi^2_1+\phi^2_3}\over{\sqrt{\phi^2}}} & {{\phi_2}\over
  {\sqrt{\phi^2}}}\cr
  -{\phi_{1}\over{\sqrt{\phi^2_1 + \phi^2_3}}}&
  {{\phi_2\phi_3}\over{\sqrt{\phi^2(\phi^2_1 + \phi^2_3})}}&
  {{\phi_3}\over {\sqrt{\phi^2}}}} \right] ;\quad {\rm with}\quad OO^T =O^TO
  =I, \quad {\rm and} \quad
  \det(O) =1.
\end{equation}
Consequently the manifestly covariant form,
\begin{equation}
S^-_{ab}= -\phi_a\phi_b \qquad {\rm with} \quad \phi^2 =\lambda,
\label{S}
\end{equation}
is a gauge equivalent description of all the Abelian instanton
configurations discussed previously. It also leads to $\phi^a
F^-_a = -\lambda \phi^a\Sigma^-_a$. $S^-$ has eigenvalues
$\{0,0,-\phi^2 = -\lambda\}$, and satisfies the contraints of
General Relativity by being symmetric and with trace $-\lambda$.

Let us further check that the symmetry reduction condition,
Eq.(\ref{Higgs}), is indeed true. It is worthwhile to recall that
Eq.(\ref{Higgs}) is gauge and diffeomorphism invariant, so it
suffices to verify that the equation holds for a particular gauge
of convenience. In the U-gauge, $S^-={\rm diag}\{0,0,
-\lambda\}$. Through $ F^-_a = S^-_{ab}\Sigma^{-b}$ we obtain
$F^-_{1,2} =0$ and $F^-_3 =-\lambda\Sigma^-_3 \neq 0$. These are
solved by $A^-_{1,2}=0$ and $F^-_3 = dA^-_3 =
-\lambda\Sigma^-_3$. With $A^-_a = (0,0,A^-_3)$, the condition
$D_{A^-}\phi_a =0$ reduces in component form to
\begin{eqnarray}
  d\phi_1 -A^-_3\wedge\phi_2&=& 0\cr
  \nonumber\\
  d\phi_2 + A^-_3\wedge\phi_1 &=& 0\cr
  \nonumber\\
  d\phi_3 &=& 0.
\end{eqnarray}
These equations are clearly satisfied by $\phi_a
=(0,0,-\sqrt{\lambda})$ in the U-gauge, and hence for all gauges
and coordinate choices. It is thus correct to associate Einstein
manifolds described by such Abelian instantons as bundle
reductions of $SO(3)$ to $U(1)$, and $S^-$ as the order parameter
which characterize them by Eq.(\ref{S}). Note that with $S^-_{ab}
=-\phi_a\phi_b$ the consistency condition of the Bianchi identity,
\begin{equation}
D_{A^-}F^-_a = (D_{A^-}S)_{ab}\wedge\Sigma^{-b}
+S_{ab}(D_{A^-}\Sigma)^{-b} = 0,
\end{equation}
is ensured because $D_{A^-}\Sigma^{-a}=0$ and $D_{A^-}\phi_a =0$.

It is to be noted that $S^-$ is itself a composite of the
vierbien through $S^-_{ab}= -*\frac{1}{2}(F^-_a\wedge\Sigma^-_b)$
with $F^-_a$ being, on-shell, the anti-self-dual part of the
curvature of the torsionless spin connection. Furthermore the
Higgs field $\phi_a$ can be gauge-transformed away, apart from a
non-vanishing third component which is associated with the
cosmological constant through $\phi_3 = -\sqrt{\lambda}$.

  These Einstein manifolds with non-vanishing $\lambda$ demonstrate
that gravitational self-interactions can generate this mechanism
of bundle reduction dynamically, despite the absence of additional
fundamental scalar fields and other physical matter. As a
corollary, we note that the (anti)self-dual part of the spin
connections of Einstein manifolds can be Abelian instantons {\it
only if} the cosmological constant is non-zero, otherwise the
condition $F^-_a = S^-_{ab}\Sigma^{-b}$ cannot be simultaneously
satisfied for $Tr(S^-) = -\lambda = 0$ and $S^- = {\rm diag}(
0,0,-\lambda \neq 0)$ for Abelian configurations.

\section{Abelian instantons and K\"{a}hler-Einstein manifolds}

The Fubini-Study and Eguchi-Hanson explicit solutions displayed in
Section III are also known to be K\"{a}hler-Einstein
manifolds\cite{EH}. This can be generalized to other Abelian
solutions. When $S^-$ is gauge equivalent to ${\rm
diag}(0,0,-\lambda)$, the
   anti-self-dual part of the spin connection is described by
   an Abelian instanton $A^-_a=(0,0,A^-_3)$. It follows that
   $F^-_a = S^-_{ab}\Sigma^{-b}$ reduce to
\begin{equation}
 F^-_{1,2}=0, \qquad F^-_3 = dA^-_3 = -\lambda\Sigma^-_3.
\end{equation}
Furthermore the equations of motion $(D_{A^-}\Sigma^-)^a =0$
become
\begin{eqnarray}
d\Sigma^-_1 - A^-_3\wedge \Sigma^-_2 &=&0,\cr
\nonumber\\
d\Sigma^-_2 + A^-_3\wedge \Sigma^-_1 &=&0,\cr
\nonumber\\
d\Sigma^-_3 &=&0. \label{Kahler}
\end{eqnarray}
Let the complex one-forms $\Omega^{1,2}$ be defined as $\Omega^1
\equiv -e^0 +ie^3$ and $\Omega^2 \equiv e^1 + ie^2$.
 The metric can then be expressed as
\begin{equation}
  ds^2 = e_A e^A = g_{\cal{A}\cal{B}}\Omega^{\cal{A}}{\overline{\Omega}}^{\cal{B}},
\end{equation}
with $g_{\cal{A}\cal{B}}=\delta_{\cal{A}\cal{B}}(\cal{A},\cal{B}$
assume values 1 to 2) being explicitly Hermitian. Furthermore, the
K\"{a}hler form, which is real, can be identified to be
\begin{equation}
\Sigma^-_3 = -e^0\wedge e^3 + e^1\wedge e^2 =
\frac{i}{2}\Omega^{\cal{A}}\wedge {\overline{\Omega}}^{\cal{A}}.
\end{equation}
A metric is said to be a K\"{a}hler if the K\"{a}hler form is
closed. But it follows from Eq.(\ref{Kahler}) that this is indeed
the case. So Einstein manifolds with $A^-$ which are Abelian
instantons can be identified as K\"{a}hler-Einstein manifolds.

\acknowledgments

The research for this work has been supported in part by funds
from the National Center for Theoretical Sciences, and the
Physics Department of the National Cheng Kung University, Taiwan.
This article is an extended version of a talk presented by the
author at the {\it International Workshop on Geometric Physics}
(July 2000) held in Hsinchu, Taiwan, at the National Center for
Theoretical Sciences.

\appendix
\section{Equivalence to Einstein's field equations}
  Let us briefly show that
\begin{equation}
D_A^- \Sigma^{-a} =0
\end{equation}
and
\begin{equation}
  F^-_a = S^-_{ab}\Sigma^{-b},
\end{equation}
with
\begin{equation}
\epsilon_{0abc}S^{-bc} =0 \quad {\rm and} \quad Tr(S^-)= -\lambda
 \end{equation}
are equivalent to Einstein's field equations.

Eq.(A1) is uniquely solved by $A^-_a=
\omega_{0a}-\frac{1}{2}\epsilon_{0abc}\omega^{bc}$, from which
$F^-_a = R_{0a} -\frac{1}{2}\epsilon_{0abc}R^{bc}$, where
$R_{AB}=d\omega_{AB} + \omega_{A}\,^C \wedge \omega^C\,_B$. The
rest of Eqs.(2.4) hold on substituting $F^-_{ab} =
S^-_{ab}\Sigma^{-b}$ provided $S^-$ is symmetric and of trace
$-\lambda$. However,
\begin{eqnarray}
  F^-_a&=&\frac{1}{2}(R_{0aAB} -\frac{1}{2}\epsilon_{0abc}R^{bc}\,_{AB})e^A\wedge
  e^B\cr
  \nonumber\\
  &=& S^-_{ab}(-e^0\wedge e^b +
  \frac{1}{2}\epsilon_{0abc}e^b\wedge e^c)
\end{eqnarray}
implies
\begin{equation}
S^-_{ab}= R_{\widetilde{0a}0b}-R_{0a0b} =
R_{0a\widetilde{0b}}-R_{\widetilde{0a}\widetilde{0b}}.
\end{equation}
These lead, together with the constraint $S^-_{ab} = S^-_{ba}$, to
\begin{equation}
R_{\widetilde{0a}0b} = R_{0a\widetilde{0b}} \quad {\rm and} \quad
R_{0a0b} = R_{\widetilde{0a}\widetilde{0b}}.
\end{equation}
It can then be shown these are equivalent to
\begin{equation}
R_{\widetilde{AB}\widetilde{CD}} = R_{ABCD},
\end{equation}
and $Tr (S^-) = -\lambda$ implies the Ricci scalar
\begin{equation}
R_{AB}\,^{AB}=R=4\lambda.
\end{equation}
It is known\cite{deFelice} that Eqs.(A7-A8) are equivalent to
Einstein's equations $R_{\mu\nu} = \lambda g_{\mu\nu}$.


\end{document}